\documentclass[aps,twocolumn,superscriptaddress,showpacs,amsmath]{revtex4}

\usepackage{epsf,latexsym}
\usepackage{amssymb,amsmath}
\usepackage{times}

\newcommand{\bra}[1]{\langle #1 |}
\newcommand{\ket}[1]{| #1 \rangle}
\newcommand{\braket}[2]{\langle #1 | #2 \rangle}

\newcommand\h{{\cal H}}
\newcommand\D{{\cal D}}
\newcommand\p{{\sf p}}

\newcommand\diag{{\mbox{diag\,}}}

\newcommand{\ignore}[1]{}

\newcommand{\ra}{{\rightarrow}}

\newcommand{\be}{\begin{equation}}
\newcommand{\ee}{\end{equation}}
\newcommand{\ba}{\begin{eqnarray}}
\newcommand{\ea}{\end{eqnarray}}

\def\CC{{\rm\kern.24em \vrule width.04em height1.46ex depth-.07ex
    \kern-.30em C}}
\def\P{{\rm I\kern-.25em P}}

\def\RR{{\rm
         \vrule width.04em height1.58ex depth-.0ex
         \kern-.04em R}}

\def\bbbone{{\mathchoice {\rm 1\mskip-4mu l} {\rm 1\mskip-4mu l}
{\rm 1\mskip-4.5mu l} {\rm 1\mskip-5mu l}}}

\def\bbbc{{\mathchoice {\setbox0=\hbox{$\displaystyle\rm C$}\hbox{\hbox
to0pt{\kern0.4\wd0\vrule height0.9\ht0\hss}\box0}}
{\setbox0=\hbox{$\textstyle\rm C$}\hbox{\hbox
to0pt{\kern0.4\wd0\vrule height0.9\ht0\hss}\box0}}
{\setbox0=\hbox{$\scriptstyle\rm C$}\hbox{\hbox
to0pt{\kern0.4\wd0\vrule height0.9\ht0\hss}\box0}}
{\setbox0=\hbox{$\scriptscriptstyle\rm C$}\hbox{\hbox
to0pt{\kern0.4\wd0\vrule height0.9\ht0\hss}\box0}}}}

\def\bbbz{{\mathchoice {\hbox{$\sf\textstyle Z\kern-0.4em Z$}}
{\hbox{$\sf\textstyle Z\kern-0.4em Z$}}
{\hbox{$\sf\scriptstyle Z\kern-0.3em Z$}}
{\hbox{$\sf\scriptscriptstyle Z\kern-0.2em Z$}}}}

\newcommand{\putfig}[2]{$$\leavevmode\hbox{\epsfxsize=#2 cm
   \epsffile{#1.eps}}$$}

\begin{document}

\title{Generalized parity measurements}

\author{Radu Ionicioiu}
\email{radu.ionicioiu@hp.com}
\affiliation{Hewlett-Packard Labs, Filton Road, Stoke Gifford, Bristol BS34 8QZ, UK}

\author{Anca E.~Popescu}
\email{anca.popescu@brl.ac.uk}
\affiliation{Bristol Robotics Laboratory, Coldharbour Lane, Bristol BS16 1QD, UK}

\author{William J.~Munro}
\affiliation{Hewlett-Packard Labs, Filton Road, Stoke Gifford, Bristol BS34 8QZ, UK}

\author{Timothy P.~Spiller}
\affiliation{Hewlett-Packard Labs, Filton Road, Stoke Gifford, Bristol BS34 8QZ, UK}

\begin{abstract}
Measurements play an important role in quantum computing (QC), by either providing the nonlinearity required for two-qubit gates (linear optics QC), or by implementing a quantum algorithm using single-qubit measurements on a highly entangled initial state (cluster state QC). Parity measurements can be used as building blocks for preparing arbitrary stabilizer states, and, together with 1-qubit gates are universal for quantum computing. Here we generalize parity gates by using a higher dimensional (qudit) ancilla. This enables us to go beyond the stabilizer/graph state formalism and prepare other types of multi-particle entangled states. The generalized parity module introduced here can prepare in {\em one-shot}, heralded by the outcome of the ancilla, a large class of entangled states, including $GHZ_n$, $W_n$, Dicke states $\D_{n,k}$, and, more generally, certain sums of Dicke states, like $G_n$ states used in secret sharing. For $W_n$ states it provides an exponential gain compared to linear optics based methods.

\end{abstract}


\maketitle

\section{Introduction}

Quantum information processing (QIP) and quantum computation (QC) promise to be disruptive technologies: applications include quantum algorithms for fast factoring \cite{shor}, database search \cite{grover} and secure key distribution \cite{crypto}. However, storing and processing information on quantum systems is difficult due to decoherence, constraining state-of-the-art quantum hardware to a few qubits. This limitation implies that at present one of the best ways to use scarce quantum resources is distributed QIP over several nodes. Useful distributed tasks may be achieved even if the total number of qubits involved is less than that which can be simulated conventionally.

Measurement based quantum computing has recently attracted considerable interest as a new paradigm for QIP. This interest has been spearheaded by two different models which complement the ``standard model'' of quantum computation, the quantum network model \cite{deutsch}. The first one initiated the field of linear optics QC \cite{klm, kok}, whereas the second started the cluster state QC \cite{1wqc}.

In this context the cluster state emerged as a quintessential resource which can be constructed before, and consumed during, computation \cite{1wqc}. A core primitive used in building the cluster state is the parity gate \cite{pittman, spin_parity}. As shown previously, the parity gate \cite{ri_parity1, ri_parity2} and the related photonic module \cite{ph_module} can be used to prepare deterministically arbitrary stabilizer/graph states, hence any cluster state used as a resource in the one-way quantum computing model \cite{1wqc}. 

A standard quantum network for the parity gate uses a qubit ancilla \cite{ri_parity1}. The new feature of our work here is that we relax this constraint and instead use a qudit ancilla \footnote{An example of using a high-dimensional (qudit) ancilla in teleportation is Ref.~\cite{louis}.}. With this we show that we can prepare, heralded by the outcome of the ancilla, a large class of entangled states in one-shot, i.e., with a {\em single} application of the generalized parity module. By tuning the dimension $d$ of the ancilla with respect to the number of input qubits $n$ we obtain several known families of entangled states: $GHZ_n$, $W_n$, Dicke $\D_{n,k}$, $G_n$ and their generalization $G_{n,k}$. These states are an important resource in several QIP protocols, including teleportation \cite{teleport}, dense coding \cite{densecode}, quantum key distribution \cite{crypto}, secret sharing ($G_n, G_{n,k}$) \cite{Gn}, $1\ra 3$ telecloning \cite{murao} and open destination teleportation ($\D_{4,2}$) \cite{D4.2}.

A very appealing feature of the generalized parity module is that it can be implemented so as to prepare directly entangled states of photons. Thus the module can be used as the enabling building block in a distributed QIP network and for small scale QIP applications. The entangled states can be created with qubits that readily distribute, without any need for interconversion.

The structure of the article is the following. In Section \ref{parity} we give a brief overview of the parity gate and its use in the photonic module. In Section \ref{general} we find the solution for the generalized parity module, then construct examples of how to prepare several classes of entangled states. We conclude in Section \ref{conclusions}.

\section{Parity measurements}\label{parity}

\subsection{The parity gate: an overview}

Historically the parity gate has been used in linear optics to construct a CNOT gate \cite{pittman}, but the outcome was probabilistic. The importance of the parity gate re-emerged in the context of fermionic quantum computation with linear elements. Beenakker {\em et al.} have shown that universality can be achieved in fermionic QC if we supplement linear gates with a single ingredient: charge parity measurements \cite{spin_parity}. This result changed the prevailing wisdom that fermionic QC cannot be done with only linear elements and (single qubit) measurements \cite{knill, terhal}. In contrast, bosons have no such limitations and universality can be achieved in photonic QC with linear gates, single photon sources and photon-number discriminating detectors, as shown by Knill, Laflamme and Milburn (KLM) \cite{klm}. The difference between bosons and fermions in terms of computational power comes from the contrasting behaviour at a beam-splitter: bunching (bosons) versus antibunching (fermions). This produced a flurry of activity in both theory and implementations, with several proposals for parity measurements in various systems \cite{engel, kolli, coish, charge_parity, stace1, virmani, mao, ruskov, nemoto, munro, spiller, pittman}.

A parity gate ($P$-gate) can be viewed---in an implementation-independent manner---simply as a black box with two inputs, $x$ and $y$, and an ancilla initialized to $\ket{0}$. The gate leaves invariant the basis states $\ket{xy}$, $x,y=0,1$ and outputs the parity $\p= x\oplus y:= x+y \mod 2$ of the inputs, i.e.:
\be
\ket{xy} \ket{0} \ra \ket{xy}\ket{x \oplus y}
\label{p2}
\ee
Upon measurement, the ancilla gives a classical bit, the parity of the input state. If the input state is a superposition $\sum_{i,j}a_{ij}\ket{ij}$, the $P$-gate projects it on a subspace of eigen-parity, i.e., on $a_{00}\ket{00}+ a_{11}\ket{11}$ (for $\p=0$) or on $a_{01}\ket{01}+ a_{10}\ket{10}$ (for $\p=1$). Building on previous work from quantum optics \cite{pittman}, Beenakker {\em et al.} \cite{spin_parity} constructed a deterministic {\em quantum} CNOT gate out of two parity gates, an ancilla and post-processing, thus proving the universality of parity measurements (along with single-qubit gates).

In effect the $P$-gate is an oracle, answering the simplest possible question when presented with two (classical) inputs $x,y$: {\em Are the two inputs equal?} in translation, $\p=0 \Leftrightarrow yes$ and $\p=1 \Leftrightarrow no$. It is surprising that such a simple gate can provide universality, where single qubit measurement failed to \cite{knill, terhal}. It confirms yet again how counter-intuitive quantum mechanics is, exemplifying how {\em less is more} in the quantum world. This is to say, knowing less (the parity), we can do more (achieve universality). The key is knowing less in a quantum sense, i.e., maintaining superposition.

As can be inferred from the action (\ref{p2}), a quantum network for the $P$ gate consists of two CNOTs, coupling each input qubit once to the ancilla, followed by a measurement of the ancilla. We can extend the network to accommodate several input qubits, each coupled once (via a CNOT) to a common ancilla, which is then measured. In this case the gate gives the parity of all $n$ inputs
\be
\ket{x_1 x_2 \ldots x_n} \ket{0} \ra \ket{x_1 x_2 \ldots x_n}\ket{\p}, \ \ \ \p= \sum_i x_i \mod 2
\ee

A very nice feature of this extension of the parity gate is that each qubit interacts only once with the ancilla. The qubits can therefore be naturally of travelling form; there is no need for them to wait around and interact again with the ancilla. For example, an extended parity gate therefore proves to be a very useful tool for preparing photonic stabilizer states and can function as a stand alone {\em photonic module} \cite{ph_module}. Suppose we have an $N$-photon pulse and that each photon interacts (sequentially) with an ancilla qubit, e.g., an atom in a cavity or an NV-center in diamond \cite{nv1}, via a simple controlled interaction: the interaction flips the atom state if the photon is $\sigma^-$ polarized and does nothing if it is $\sigma^+$ polarized. For simplicity in the following we use $\ket{\pm}= (\ket{0}\pm \ket{1})/\sqrt{2}$ states which differ from $\sigma^\pm$ by a simple phase-shift, $\sigma^\pm= \diag(1,i) \ket{\pm}$. Thus we assume the interaction
\begin{eqnarray}
\ket{+}\ket{0}_a &\ra& \ket{+}\ket{0}_a \cr
\ket{-}\ket{0}_a &\ra& \ket{-} X\ket{0}_a
\end{eqnarray}
where $\ket{i}_a$ denotes the ancilla state; in the following we will denote the Pauli operators for a qubit by $X,Y,Z$. This photonic module can prepare an arbitrary $N$-photon stabilizer state using only parity measurements on the ancilla qubit (the atom) and single qubit gates. Given an $N$-photon state $\ket{\Psi}_N$ interacting sequentially with the ancilla qubit $\ket{i}_a$, the action of the photonic module is \cite{ph_module, ph_module2}:
\be
\ket{\Psi}_N \ket{i}_a \ra (P_0 \otimes \bbbone + P_1 \otimes X) \ket{\Psi}_N \ket{i}_a
\ee
where $P_k:= \frac{1}{2} [ \bbbone+ (-1)^k X^{\otimes N} ]$ are even/odd parity projectors acting on the $2^N$-dimensional photon space.

Let us now start to go beyond simple parity gates made from CNOTs acting on the ancilla. The first question we address is the following: apart from the NOT gate, are there other unitary transformations $U \in U(2)$ acting on the ancilla, such that the Control-$U$ gate (controlled by an input qubit) can be used to construct a parity gate? At first sight, there are two requirements for a good $U$. First, we need to have $\ket{\phi} \perp U\ket{\phi}$, for a suitable ancilla state $\ket{\phi}$. This is essential in order to unambiguously distinguish odd and even parity states. The second condition is $U^2= \bbbone$, as we want the states $\ket{00}$ and $\ket{11}$ of the qubits to be indistinguishable. In the next section we will find the general solution of this problem and we will show that the second requirement is not independent: it is a simple consequence of the orthogonality condition, which is the crucial one.

\subsection{Ancilla as a qubit}\label{qubitancilla}

In this section we answer the previous question and find the general unitary $U\in U(2)$ that can be used to construct a parity gate. The solution has practical consequences: it enables to find the right interactions required to implement a $P$-gate.

We first consider the case where the ancilla is still a qubit, so $d=2$. We are looking for the unitaries $U\in U(2)$ and the states $\ket{\phi}$ such that $\ket{\phi}$ and $\ket{U \phi}:= U \ket{\phi}$ are orthogonal, namely
\be
\braket{\phi}{U \phi}= 0
\label{orto1}
\ee

By diagonalizing $U$, $U= V D V^\dag$, with $V\in U(2)$ and $D= e^{i\varphi_0} \diag(1, e^{i\varphi_1})$, the previous problem is equivalent to finding $\ket{\psi}:= V^\dag \ket{\phi}$ satisfying $\braket{\psi}{D \psi}= \braket{\phi}{U \phi}= 0$. Clearly $V$ is just a change of basis, so the physics is in the eigenvalues. Neglecting the overall phase $e^{i\varphi_0}$, the solution follows immediately
\begin{eqnarray}
\label{Dsol}
D&=& \begin{pmatrix} 1 & \cr & -1 \end{pmatrix}= Z\cr
\ket{\psi}&=& (\ket{0}+ e^{i\varphi} \ket{1})/\sqrt{2}
\end{eqnarray}

Now although very simple, this solution has an intuitive geometric interpretation which will provide inspiration for the generalized solution that forms the main result of this paper. Furthermore, the key features of the result can be expressed mathematically in a form that lends naturally to generalization.

First, we observe that the states $\ket{\psi}$ are on the equator of the Bloch sphere and thus are perpendicular to the $Oz$ axis associated with $D=Z$. Now any unitary $U$ can be viewed as a rotation of the Bloch sphere through angle $\alpha$ around an axis $\vec n$, $U= \exp(i\alpha \vec \sigma . \vec n)$ (with $\vec \sigma=(X,Y,Z)$). Therefore, in general, the states $\ket{\phi}$ satisfying $\braket{\phi}{U \phi}=0$ are on the great circle of the Bloch sphere perpendicular to $\vec n$. Moreover, as the solution (\ref{Dsol}) satisfies $D^2= \bbbone$, we obtain $U^2= \bbbone$ rather than imposing it, and so $\alpha= \pi/2$. Hence $U$ is of the form $U= i\vec \sigma . \vec n$.

Second, we observe that the equator of the Bloch sphere is the orbit of $\ket{+}$ under the action of the group $G= \{ \diag(1, e^{i\varphi}) \}$. Thus all the states $\ket{\psi}$ satisfying $\bra{\psi} Z \ket{\psi}=0$ can be written as $\ket{\psi}= g\ket{+}$, with $g\in G$. Note also that the group $G$ is nothing but the commutant of $Z$ (up to a global phase), namely $Z':= \{ M\in U(2), [M, Z]=0 \} = \{ \diag(e^{i\theta_0}, e^{i\theta_1}) \}$.

\section{Generalized parity}\label{general}

We are now ready to relax the constraint of the qubit ancilla and explore the general case where we have at our disposition a higher dimensional space, i.e., a qudit. Before proving the general result it will be illuminating to see an example.

\subsection{A simple application: $W$-states}

\begin{figure}
\putfig{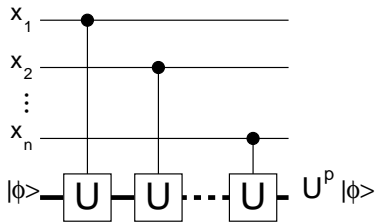}{5}
\caption{A generalized parity module. The generalized parity is defined as $\p= \sum_i x_i \mod d$, with $x_i=0,1$. The dimension of the ancilla (bold line) Hilbert space is $\dim \h= d$. If the ancilla is a qubit $d=2$ and $U=X$, the network is equivalent to the photonic module discussed in \cite{ph_module}.}
\label{gen_parity}
\end{figure}

In the simplest generalization of the $P$ gate we have three qubits coupled to a common qutrit ancilla, as in Fig.~\ref{gen_parity} with $d= n= 3$. In this case we are looking for a unitary $U\in U(3)$ and a vector $\ket{\psi}$ such that the set $\{ \ket{\psi}, U\ket{\psi}, U^2\ket{\psi} \}$ is orthonormal. A particular solution is given by
\begin{eqnarray}
U&=& \diag(1,\omega, \omega^2) \cr
\ket{\psi}&=& (\ket{0}+ \ket{1}+ \ket{2})/\sqrt{3}
\label{Wsol}
\end{eqnarray}
with $\omega=e^{2\pi i/3}$. Using the identities $1+\omega+\omega^2=0$ and $U^3=\bbbone$, it can be easily shown that the above solution (\ref{Wsol}) satisfies the required orthogonality conditions.

A natural question arises: {\em What is this useful for?} We show that the simple network in Fig.~\ref{gen_parity} for the case $d=3$ can prepare (probabilistically) $W$-states,
\be
\ket{W}= (\ket{001}+ \ket{010}+ \ket{100})/\sqrt{3}.
\ee
Suppose that the initial product state of the qubits is the equal superposition of all basis states, i.e., $\ket{+}^{\otimes 3}= 2^{-3/2} \sum_{i=0}^7 \ket{i}$. The ancilla qutrit is prepared in the initial state $\ket{\psi}$ and, after interacting with the three qubits, is measured. Since the three possible states of the ancilla are orthogonal, they can be distinguish with certainty and upon the projective measurement, the qubit register is in one of the three possible states:
\begin{eqnarray}
\ket{GHZ}&=& (\ket{000}+ \ket{111})/\sqrt{2} \ \ , \ \ p= 1/4 \cr
\ket{W}&=& (\ket{001}+ \ket{010}+ \ket{100})/\sqrt{3} \ \ , \ \ p= 3/8 \cr
\ket{W'}&=& (\ket{110}+ \ket{101}+ \ket{011})/\sqrt{3}, \ p= 3/8
\end{eqnarray}
where $p$ is the probability. As $\ket{W'}= X^{\otimes 3} \ket{W}$, this simple quantum circuit prepares $W$-states with probability $p(W)=3/4$.  Post-processing, i.e., locally bit flipping all qubits, can be applied to transform $\ket{W'}$ to $\ket{W}$ if required; alternatively this classical information can be supplied along with the $W$-state, dependent upon what it is to be used for. The best method so far for producing $W$ states using linear elements and post-selection  has a probability of success of $3/16$ \cite{tashima}, hence four times lower.

It is worth emphasising that $W$-states are {\em not} stabilizer/graph states, hence they cannot be described in the stabilizer formalism. As such they cannot be prepared systematically, in one-shot, using the photonic module or $P$-gates \cite{ri_parity1, ph_module} as described above (they require extra resources/ancill\ae). It is known that $W$-states belong to a different entanglement class than $GHZ$-states, and the two families cannot be interconverted through local operations and classical communications (LOCC) \cite{w_states}; thus they represent different entanglement resources. For example, $W$-states are more robust under qubit losses than $GHZ$-states.

\subsection{Ancilla as a qudit}

We are now ready to prove the general result, addressing the case where the ancilla is a qudit, so $\dim \h= d$. We let $\{ \ket{0}, \ldots, \ket{d-1} \}$ be the computational (or $Z$) basis in $\h$. Let $\bbbz_d= \{ 0, 1, \ldots, d-1\}$. The generalized Pauli operators $X_d$ and $Z_d$ for qudits are:
\begin{eqnarray}
X_d \ket{i}&=& \ket{i\oplus 1} \cr
Z_d \ket{i}&=& \omega^i \ket{i}
\label{xz}
\end{eqnarray}
with $\omega:= e^{2\pi i/d}$ and $\oplus$ now addition mod $d$. Thus $Z_d:= \diag(1, \omega, \omega^2, \ldots, \omega^{d-1})$ in this basis. We will also need the Fourier (or $X$) basis, defined as the Fourier transform of the $Z$ basis:
\begin{eqnarray}
\ket{u_k}&=& d^{-1/2} \sum_{j=0}^{d-1}\ \omega^{-kj} \ket{j} \cr
\ket{j}&=& d^{-1/2} \sum_{k=0}^{d-1}\ \omega^{jk} \ket{u_k}
\end{eqnarray}
from which follows the useful identity $d^{-1} \sum_j \omega^{jk}= \delta_{0k}$. The action of the Pauli operators on this basis is:
\begin{eqnarray}
X_d \ket{u_k}&=& \omega^k \ket{u_k} \cr
Z_d \ket{u_k}&=& \ket{u_{k-1}}
\end{eqnarray}

Generalizing the problem of section \ref{qubitancilla} to the full problem, we now want to find a unitary $U\in U(d)$ and a state $\ket{\phi}\in \h$ such that the set $\{ \ket{\phi}, U\ket{\phi}, ..., U^{d-1} \ket{\phi} \}$ is orthonormal; this is necessary as we want to discriminate unambiguously between states of different parity. The analogue of the previous condition (\ref{orto1}) is now:
\be
\braket{\phi}{U^i \phi}= 0,\ \ \ \ \forall i=1..d-1.
\ee
Following similar reasoning to that of section \ref{qubitancilla}, we need only to consider diagonal unitaries. Let $U= V D V^\dag$, with $D=\diag(\lambda_0,..,\lambda_{d-1})$ containing the eigenvalues $\lambda_i:= e^{i\varphi_i}$ of $U$; then $\braket{\phi}{U\phi}= \braket{\psi}{D\psi}=0$ with $\ket{\psi}:= V^\dag \ket{\phi}$. Therefore, we can focus on diagonal unitaries $D \in U(d)$ satisfying
\be
\braket{\psi}{D^i \psi}= \delta_{0i},\ \ \ \ \forall i \in \bbbz_d
\label{orto_d}
\ee
for some $\ket{\psi}\in \h$. The objective is to find the solutions analogous to those given in (\ref{Dsol}).

First, we observe that it is clear that not all $U$ will have solutions to the above equation. Indeed, suppose the unitary is an infinitesimal rotation, $U= \exp(i\epsilon Z)$, $\epsilon \approx 0$, i.e., very close to the identity. Then $\braket{\phi}{U\phi}\simeq 1+ i\epsilon \bra{\phi} Z \ket{\phi} \approx 1$ for all $\ket{\phi}\in \h$.

Second, generalizing the properties of the states given in (\ref{Dsol}), we observe that if for a given $D$ there is a vector satisfying $\braket{\psi_0}{D^i \psi_0}= 0$, then the orbit of $\ket{\psi_0}$ under the action of the commutant $D':= \{ g\in U(d), [g, D]=0 \}$ will also be a solution. Hence, if $\ket{\psi'}= g\ket{\psi_0}$ with $g\in D'$, then $\braket{\psi'}{D^i \psi'}= \braket{\psi_0}{g^\dag D^i g\psi_0}= \braket{\psi_0}{D^i \psi_0}= 0$.

We decompose the state as $\ket{\psi}= \sum_{j\in \bbbz_d} a_j \ket{j}$, so equation (\ref{orto_d}) becomes
\be
\sum_{j=0}^{d-1} |a_j|^2 \lambda_j^i = \delta_{0i}, \ \ \forall i \in \bbbz_d.
\label{syseq}
\ee
There are two possible cases to address, dependent upon the nature of the eigenvalues of $D$.

{\em Non-degenerate case.} We assume all the eigenvalues of $D= \diag(\lambda_0, \ldots, \lambda_{d-1})$ are distinct, $\lambda_j \ne \lambda_k,\ \ \forall j\ne k$. Using the expansion of the Vandermonde determinant, we obtain
\begin{eqnarray}
\label{aj2}
\frac{1}{|a_j|^2} &=& \prod_{i\in \bbbz_d,\ i\ne j} \left( 1-\frac{\lambda_j}{\lambda_i} \right) \cr
&=&  \prod_{i\in \bbbz_d,\ i\ne j} 2 \left| \sin\frac{\varphi_j-\varphi_i}{2} \right| e^{\frac{\varphi_j- \varphi_i}{2}- \frac{\pi}{2}+ n_{ij}\pi}
\end{eqnarray}
with $n_{ij} \in \bbbz$; the last equation follows from the polar decomposition of each term in the product. Requiring the right hand side to be real and positive, it follows that $\varphi_j= \varphi_0+ \frac{2\pi}{d} m_j$, for some integers $m_j$. Since all the phases are different (as the eigenvalues are non-degenerate), we can take $\varphi_j= \varphi_0+ \frac{2\pi}{d} j$, modulo a reordering of the eigenvalues. The unitary transformation we are looking for is
\be
D= e^{i\varphi_0} \diag(1, \omega, \omega^2, \ldots, \omega^{d-1}) = e^{i\varphi_0} Z_d.
\ee
Substituting these phases into eq.~(\ref{aj2}) we get
\be
\frac{1}{|a_j|^2}= \prod_{i\in \bbbz_d,\ i\ne j} (1- \omega^{j-i})= \prod_{1\le i \le d-1} (1- \omega^i)= d
\label{aj2d}
\ee
This shows that $\ket{\psi}$ is an equal superposition of all basis states
\be
\ket{\psi}= d^{-1/2} \sum_i e^{\theta_i} \ket{i}
\ee
As before, there is an appealing geometric interpretation. The entire manifold of solutions ${\cal M}= \{ \ket{\psi} \}$ can be generated by acting with the commutant of $Z_d$ (in $U(d)$) on a single state, say $\ket{u_0}= d^{-1/2} \sum_i \ket{i}$; hence ${\cal M}= Z'_d \ket{u_0}$, i.e., is the orbit of $\ket{u_0}$ under $Z'_d$. The commutant is the set $Z'_d= \{ \diag(e^{\theta_0},..., e^{\theta_{d-1}}) \}$. Thus we have proved the following:

{\bf Proposition:} Let $\ket{\phi}\in \h$, $\dim \h=d$, and $U\in U(d)$ a unitary acting on $\h$ such that the set $\{ \ket{\phi}, U\ket{\phi}, ..., U^{d-1} \ket{\phi} \}$ is orthonormal. If $U$ has nondegenerate eigenvalues, then
\begin{eqnarray}
U&=& V Z_d V^\dag \cr
\ket{\phi}&=& V g \ket{u_0}
\end{eqnarray}
where $V\in U(d)$ is an arbitrary unitary and $g\in Z'_d$ belongs to the commutant of $Z_d$.

{\em Degenerate case.} We assume that $D$ has degenerate eigenvalues, $D=\diag(\lambda_0 \bbbone_{k_0}, \ldots, \lambda_{s-1} \bbbone_{k_{s-1}})$, where $k_i$'s are the degeneracies of the $s$ distinct eigenvalues and $\sum_{i \in \bbbz_s} k_i= d$. In this case the system (\ref{syseq}) is singular and we have only $s$ independent equations with a Vandermonde discriminant. As before, the eigenvalues are $\lambda_j= e^{i \phi_0} \omega^j$, with $\omega= e^{2\pi i/s}$ a root of unity of degree $s$. Since $D^s= \bbbone$, now we can have only $s$ orthonormal vectors $\{ \ket{\psi}, D\ket{\psi}, .., D^{s-1} \ket{\psi} \}$. The absolute value of the amplitudes $|a_i|$ are no longer fixed as in the nondegenerate case from eq.~(\ref{aj2d}); in this case we have only $s$ constraints for $d$ variables $|a_j|^2$, namely
\begin{eqnarray}
|a_0|^2 &&+ \ldots +|a_{k_0-1}|^2= 1/s \cr
&\vdots& \cr
|a_{d-k_s}|^2 &&+ \ldots + |a_{d-1}|^2 =1/s
\end{eqnarray}
where now all the amplitudes $|a_i|$ belonging to a degenerate eigenspace are on a hypersphere of radius $s^{-1/2}$, generalizing eq.~(\ref{aj2d}).

\subsection{A generalized parity module}

Having determined the general solution for $U$, we can now calculate the action of the generalized parity module (Fig.~\ref{gen_parity}) on an arbitrary state of the qubits. We discuss two cases, namely $U=Z_d$ and $U=X_d$. In order to make the connection with the photonic module \cite{ph_module}, we assume the qubits are photons interacting with an atom in a cavity (the qudit ancilla). Of course, this is not the only possible implementation, but we use it here as an illustration of the application of the generalized parity module.

We assume the action of the module on a single photon qubit is
\begin{eqnarray}
\ket{0} \ket{\phi} &\ra& \ket{0} \ket{\phi} \cr
\ket{1} \ket{\phi} &\ra& \ket{1} Z_d \ket{\phi}
\label{0z}
\end{eqnarray}
Now, if the atomic ancilla is in one of the $Z$-basis states $\ket{j}$, the transformation of an arbitrary photon state $\ket{\psi}= a\ket{0}+ b\ket{1}$ is \cite{ph_module2}
$$\ket{\psi} \ket{j} \ra [a\ket{0}+ \omega^j b\ket{1}] \ket{j}= [A_1^j \ket{\psi}] \ket{j}$$
where now the photon gets a phase shift $A_1^j= \diag(1, \omega^j)$ dependent upon the basis state $\ket{j}$. The action of the module on a general $N$-photon state $\ket{\Psi}_N$ follows straightforwardly, as each photon interacts independently with the module: $\ket{\Psi}_N \ket{j} \ra [A_N^j \ket{\Psi}_N] \ket{j}$; here $A_N:= A_1^{\otimes N}$ is a tensor product of identical single-qubit phase shifts acting on each photon. If, alternatively, the ancilla is in one of the $X$-basis states $\ket{u_k}$, we have:
\be
\ket{\Psi}_N \ket{u_k} \ra \sum_{i=0}^{d-1} P_i \otimes Z_d^i \ \ket{\Psi}_N \ket{u_k}
\ee
The projectors are defined as
\be
P_i := d^{-1} \sum_k \omega^{-ik} A_N^k
\label{P_i}
\ee

It is easy to see that the operators $\{ P_j \}$ have the following properties:\\
(i) $P_j^\dag= P_j$;\\
(ii) $P_j P_k= \delta_{jk} P_k$;\\
(iii) $\sum_j P_j =\bbbone$,\\
hence they form a complete set of orthogonal projectors.

For a state $\ket{\Psi}$, the probability of projecting on the $j$ parity subspace is
\be
p(j)= \bra{\Psi} P_j \ket{\Psi}
\ee
The dimension of the $j$-th parity subspace is $\dim P_j= 2^N \bra{+}^N P_j \ket{+}^N$, where $\ket{+}^N:= H^{\otimes N} \ket{0}^{\otimes N}$ is the equal superposition state of $N$ qubits and $H$ is the Hadamard gate. Since the $P_j$'s are a complete set of projectors, we obviously have $\sum_j \dim P_j= 2^N$.

The interaction (\ref{0z}) is suitable if the photon encodes a dual-rail (or mode) qubit and the ancilla (e.g., an atom in a cavity) is situated in rail 1, in which case a $Z_d$ gate is enacted on the ancilla. However, as discussed in section \ref{parity}, for some systems a more natural interaction is with the $\sigma^\pm$ polarization states of the photon. Neglecting a trivial phase, we therefore also consider the following interaction
\begin{eqnarray}
\ket{+}\ket{\phi} &\ra& \ket{+}\ket{\phi} \cr
\ket{-}\ket{\phi} &\ra& \ket{-} X_d \ket{\phi}.
\end{eqnarray}
Then the action of the module is given by (with the ancilla in the $Z$-basis state $\ket{j}$)
\be
\ket{\Psi}_N \ket{j} \ra \sum_{i=0}^{d-1} \tilde P_i \otimes X_d^i \ \ket{\Psi}_N \ket{j}
\ee
The new projectors are $\tilde P_i = d^{-1} \sum_k \omega^{-ik} B_N^k$, with $B_N:= B_1^{\otimes N}$; $B_1= H \diag(1, \omega) H= H A_1 H$ is a single qubit $x$-rotation on the photon. Mathematically, this is nothing but a change of basis compared to (\ref{0z})-(\ref{P_i}), but this is relevant from a physical perspective: given a quantum system, certain gates are easier to implement experimentally than others. For example, atoms interact naturally with circularly polarized light and the same holds for excitons created in quantum dots.

It is worth mentioning an interesting duality property between the two actions discussed above: $Z_d$ acts on (the ancilla prepared in) $\ket{u_0}$, which is the Fourier transform of the $Z$-basis state vector $\ket{0}$; similarly, $X_d$ acts on the vector $\ket{0}$, which is the (inverse) Fourier transform of its own basis eigenvector $\ket{u_0}$. In other words, $Z_d$ acts on the eigenvectors of its Fourier transform $X_d$ (and vice-versa).

\subsection{Preparation of Dicke states}

Having calculated the action of the generalized parity module on arbitrary input states of qubits, we use these results to show how the module can prepare certain classes of quantum states, interesting from the perspective of quantum information and/or many-body physics. Our first example is the class of Dicke states \cite{dicke}. These are symmetric states of $n$ particles with $k$ excitations (i.e., 1's) and can be seen as multiparticle generalizations of $W_n$ states \cite{dicke, toth}:
\be
\D_{n,k}= {n \choose k}^{-1/2} \sum_j {\cal S}_j^{(n)} \ket{0}^{\otimes n-k} \ket{1}^{\otimes k}
\ee
where the sum is over all {\em distinct} permutations ${\cal S}_j^{(n)}$ of $n$ particles. Example of Dicke states are $n$-particle $W_n$ states, $W_n= \D_{n,1}= n^{-1/2}(\ket{10..0}+ .. +\ket{00..1})$. They satisfy a simple duality property:
\be
\D_{n,n-k}= X^n \D_{n,k}
\label{dual}
\ee
where $X^n:= X^{\otimes n}$ is a bit flip on all qubits.

The generalized parity module can prepare Dicke states in one shot, i.e., with a single application of the module. In any one run, the exact state prepared is heralded by the measurement outcome of the ancilla. Consider the case where the dimension of the ancilla space is $d=n$, i.e., equal to the number of qubits. Assume the initial product state of the $n$ qubits is $\ket{+}^{\otimes n}= 2^{-n/2}\sum_{j\in \bbbz^n} \ket{j}$, an equal superposition of all basis states. Applying the parity module (Fig.~\ref{gen_parity}) on this state will project it to one of the following $n$ states:
\be
\{ GHZ_n, \D_{n,1}=W_n,..., \D_{n,k},..., \D_{n,n-1}= X^n W_n \}
\ee
Taking into account the duality property (\ref{dual}), for $n>2$ there are only $\lfloor n/2 \rfloor +1$ distinct states (up to local bit flips). If $n=2$ there is only one distinct state, since $\ket{00}+\ket{11}$ and $\ket{01}+\ket{10}$ are locally equivalent. The probability of obtaining one of these states is
\begin{eqnarray}
p(GHZ_n)&=& 2^{-n+1} \cr
p(\D_{n,k})&=& 2^{-n+1} {n \choose k}
\label{prob}
\end{eqnarray}
The probability peaks for Dicke states having half the number of excitations $\D_{n,n/2}$. For this specific example the probability scales extremely well, only damping with the root of the qubit number, so $p(\D_{2k,k}) \sim \frac{2}{\sqrt{\pi k}}$.

How efficient is this method compared to other means of preparing $W_n$ states? From (\ref{prob}) we have $p(W_n)= n 2^{1-n}$. In a recent article \cite{tashima} the success probability for producing $W_n$ states using linear elements and post-selection was $p(W_n)= n 2^{2-2n}$ ($n$ odd) and $n 2^{3-2n}$ ($n$ even). This shows that our method gives an exponential gain of at least $2^{n-2}$ compared to the method in Ref.~\cite{tashima}. 

Before going further, we will review briefly the importance of these states. From a theoretical point of view various Dicke states have different entanglement properties and as such it is important to understand and characterize them.  A recent study \cite{guhne} showed that $\D_{4,2}$ states are more robust under decoherence that $W_4$, $GHZ_4$ and linear cluster states $CL_4$. Also, $W_n$ states lead to stronger nonclassicality than $GHZ_n$ states \cite{sen}. $\D_{4,2}$ can be used in $1\ra 3$ telecloning and open destination teleportation \cite{D4.2}; it has another interesting property: measuring one qubit, one can obtain either a $W_3$ or a $GHZ_3$ state. As mentioned before, these two states belong to different entanglement families and cannot be transformed into each other by stochastic local operations and classical communication.

Several of these states have been observed experimentally in various systems. These include ion traps ($W_4,...,W_8$ \cite{W8} and $GHZ_6$ \cite{GHZ6}) and photons ($\D_{4,2}$ \cite{D4.2}).

\subsection{The case $n>d$}

In the previous section the number of qubits was equal to the dimension of the ancilla. Another interesting case is $n>d$. ($n<d$ is trivial.)

Suppose we again prepare the qubits in the equal superposition state $\ket{+}^{\otimes n}$. If after the measurement the ancilla is found to be $k$, $k=0,...,d-1$, then the qubits are projected to (in the following we neglect normalizations):
\begin{eqnarray}
\psi_k&=& \sum_{x:\, \p(x)=k \mod d} \ket{x}= {n \choose k}^{1/2} \D_{n,k} \cr
&+& {n \choose k+d}^{1/2} \D_{n, k+d}+ \ldots
\label{psi_k}
\end{eqnarray}
where $\D_{n,0}= \ket{0}^{\otimes n}$. The sum is over all basis states of $n$ qubits $\ket{x}:=\ket{x_1 x_2...x_n}$ such that the number of 1's is $k \mod d$, $\p(x)= \sum_j x_j= k \mod d$. Thus $\psi_k$ is a weighted sum of Dicke states, and in general there is no simple way of characterizing such sums.

It is insightful to analyse a few examples and see how the projected states vary, first, with the dimension $d$ of the ancilla (at a fixed number of qubits $n$) and second, with increasing number of qubits (when the ancilla has the same dimension).

{\em Example 1:} $n=4, d=3$. Upon measurement of the ancilla, we obtain one of the following states (the subscript indicates the eigenvalue of the measured ancilla, i.e., the generalized parity):
\begin{eqnarray}
\psi_0 &=& 0000+ 1110+ 1101+ 1011+ 0111= X^4 \psi_1 \cr
\psi_1 &=& 0001+ 0010+ 0100+ 1000+ 1111= 2 W_4+ 1111 \cr
\psi_2 &=& \D_{4,2}
\end{eqnarray}

{\em Example 2:} $n=5, d=3$. Increasing by one the number of qubits but keeping the ancilla the same we obtain (again, up to normalization):
\begin{eqnarray}
\psi_0 &=& 00000+ \sqrt{10}\D_{5,3} \cr
\psi_1 &=& \D_{5,1}+ \D_{5,4}= (\bbbone+ X^5) W_5 \cr
\psi_2 &=& \sqrt{10}\D_{5,2}+ 11111= X^5 \psi_0
\end{eqnarray}

{\em Example 3:} $n=5, d=4$. In this case the four projected states are:
\begin{eqnarray}
\psi_0 &=& 00000+ \sqrt{5}\D_{5,4}= X^5 \psi_1 \cr
\psi_1 &=& \sqrt{5}W_5+ 11111 \cr
\psi_2 &=& \D_{5,2} \cr
\psi_3 &=& \D_{5,3}= X^5 \psi_2
\end{eqnarray}

\subsection{Generalized $G_n$ states and secret sharing}

An interesting family of states is $G_n$ introduced in \cite{Gn}
\be
G_n:= \frac{1}{\sqrt 2}(W_n+ X^n W_n)
\ee

{\em Example 4:} $d=n-2$. In this case one of the outcomes of the parity module are $G_n$ states (we omit normalization):
\begin{eqnarray}
\psi_0 &=& 0^{\otimes n}+ {n \choose 2}^{1/2} \D_{n,n-2}= X^n \psi_2 \cr
\psi_1 &=& W_n+ \D_{n,n-1}= (\bbbone+ X^n) W_n= G_n \cr
\psi_2 &=& {n \choose 2}^{1/2} \D_{n,2}+ 1^{\otimes n} \cr
\vdots \cr
\psi_{n-3} &=& \D_{n,n-3}= X^n \D_{n,3}
\end{eqnarray}
As shown in Ref.~\cite{Gn}, the $G_n$ states can be used for secret sharing. In this protocol, Alice (the secret holder) wants to distribute her secret among $n-1$ parties $B_1,\ldots, B_{n-1}$ (the Bobs) such that all these Bobs have to cooperate to find out the secret; hence, if at least one Bob is left outside, the remaining ones cannot recover Alice's secret.

Define the following generalization of $G_n$ states:
\begin{eqnarray}
G_{n,k}&:=& \frac{1}{\sqrt{2}} (\D_{n,k}+ X^n \D_{n,k}), \ \ n\ne 2k \cr
G_{2k, k}&:=& \D_{2k, k}
\end{eqnarray}
since $X^n \D_{2k, k}= \D_{2k, k}$; we obviously have $G_n= G_{n,1}$.

From eq.~(\ref{psi_k}) we notice that if $k+d=n-k$ and $k<d$, the $n$-qubit state corresponding to the $k$-th value of the ancilla is $\psi_k= \frac{1}{\sqrt{2}}(\D_{n,k}+ \D_{n,n-k})= G_{n,k}$, so we have the following:

{\em Example 5:} $d= n-2k, k<d$. The parity module can naturally prepare $G_{n,k}$ states heralded by the $k$-th value of the ancilla.

A simple calculation shows:
\begin{eqnarray}
\bra{G_{n,k}} X^n \ket{G_{n,k}} &=& 1 \cr
\bra{G_{n,k}} Y^n \ket{G_{n,k}} &=& \begin{cases} 0, \ \ n=2m+1 \cr (-1)^{m+k}, \ \ n=2m \end{cases}
\end{eqnarray}
and obviously $\bra{G_{2k,k}} Y^n \ket{G_{2k,k}}=1$. The previous properties are analogous to those of the $G_n=G_{n,1}$ states, which are essential for secret sharing \cite{Gn}. Using a similar argument as in Ref.~\cite{Gn}, we conjecture that the $G_{2m,k}$ states can also be used for secret sharing.

All these examples demonstrate the flexibility of the generalized parity module in preparing various forms of interesting and potentially useful entangled states, by varying the dimension of the ancilla and the number of qubits. Other states can be obtained if we use an initial state different from $\ket{+}^{\otimes n}$.

Although the success probability for some, but not for all, of the states decreases exponentially with $n$, as in eq.~(\ref{prob}), the main advantage of the generalized parity module is that it can prepare---heralded and in a single shot---a large spectrum of different families of entangled states: $GHZ_n$, $W_n$, $D_{n,k}$, $G_n$ and generalized $G_{n,k}$ states. We are not aware of any unified protocol or quantum gate which can prepare such a diverse set of useful entangled states with relatively simple resources: a qudit ancilla which interacts, sequentially and homogeneously, with $n$ qubits. Our method is certainly useful to prepare various entangled states, using modest qubit and ancilla resources. For relatively small numbers of qubits---the likely experimental situation in the near future---the exponential damping (with qubit number) of specific preparation probabilities is not really an issue. Our approach therefore offers a very flexible tool for the future laboratory preparation of a wide range of entangled states.

\section{Conclusions}\label{conclusions}

One of the major breakthroughs in quantum information was the insight that measurements are not only useful as the final step of a computation, but can be used during the computation itself. The KLM model \cite{klm} initiated the field of linear optics QC by proving that photon discriminating detectors and active feed-forward can provide the nonlinearity required for a photonic two-qubit gate. On the other hand, in cluster state QC \cite{1wqc} any quantum algorithm can be performed using single-qubit measurements (plus feed-forward) performed on a highly entangled initial state; thus the cluster state and single qubit measurements are universal resources for QC.

Standard resources for preparing stabilizer and cluster states are parity gates \cite{ri_parity1} and photonic modules \cite{ph_module, ph_module2}. In this article we introduced a generalized parity module and studied its use in preparing several families of entangled states. We have shown that using a qudit ancilla we can produce in one-shot measurements a large class of multiparticle entangled states, heralded by the measurement outcome of the ancilla. It is somehow surprising that such a simple circuit can prepare a large class of entangled states, like $GHZ_n$, $W_n$, Dicke $\D_{n,k}$ and $G_{n,k}$, with the number of qubits $n$ and the dimension of the ancilla $d$ as the only free parameters. For $W_n$ states, our model provides an exponential gain compared to linear optics and post-selection \cite{tashima}. The previous states are essential in several quantum information protocols; examples include teleportation, dense coding, quantum key distribution, secret sharing ($G_n, G_{n,k}$), $1\ra 3$ telecloning and open destination teleportation ($\D_{4,2}$).

An important feature of the parity module is that all qubits interact once only and in the same way with the ancilla. This is particularly relevant in the case of the photonic module \cite{ph_module, ph_module2}, for example. The qubits (photons) are sent sequentially through a cavity containing an atom (or a QD in a photonic crystal); the cavity is prepared in a known state and subsequently measured after interacting with all photons. This means that there is no need of extra pulses applied to the cavity between the photons, resulting in a simplified design. Furthermore, the resultant entanglement is between photons, which are naturally amenable to distribution. A straightforward application of these highly entangled states is therefore in a distributed QIP network. Such states could enable useful quantum tasks, even with very modest numbers of qubits.

\acknowledgments

We thank European Union for support through the QAP project.

%%%%%%%%%%%%%%%%%%%%%%%%%%%%%%%%%%%%%%%%%%%%


\begin{thebibliography}{}

\bibitem{shor} P.W.~Shor, SIAM J.~Sci.~Statist.~Comput.~{\bf 26}, 1484 (1997).

\bibitem{grover} L.~Grover, \prl  {\bf 79}, 325 (1997).

\bibitem{crypto} N.~Gisin, G.~Ribordy, W.~Tittel, and H.~Zbinden, \rmp {\bf 74}, 145 (2002).

\bibitem{deutsch} D.~Deutsch, Proc.~Roy.~Soc.~Lon.~A {\bf 425}, 73 (1989).

\bibitem{klm} E.~Knill, R.~Laflamme, and G.~Milburn, Nature {\bf 409}, 46 (2001).

\bibitem{kok} P.~Kok, W.J.~Munro, K.~Nemoto, T.C.~Ralph, J.P.~Dowling, G.J.~Milburn, \rmp {\bf 79}, 135 (2007).

\bibitem{1wqc} R.~Raussendorf and H.J.~Briegel, \prl {\bf 86}, 5188 (2001).

\bibitem{pittman} T.B.~Pittman,B.C.~Jacobs, and J.D.~Franson, \pra {\bf 64}, 062311 (2001).

\bibitem{spin_parity} C.W.J.~Beenakker, D.P.~DiVincenzo, C.~Emary, and M.~Kindermann, \prl {\bf 93}, 020501 (2004); quant-ph/0401066.

\bibitem{ri_parity1} R.~Ionicioiu, \pra {\bf 75}, 032339 (2007).

\bibitem{ri_parity2} R.~Ionicioiu, Int.~J.~Quant.~Inform., {\bf 5}, 3 (2007).

\bibitem{ph_module} S.J.~Devitt, A.D.~Greentree, R.Ionicioiu, J.L.~O'Brien, W.J.~Munro, L.C.L.~Hollenberg, \pra {\bf 76}, 052312 (2007).

\bibitem{louis} S.G.R.~Louis, A.D.~Greentree, W.J.~Munro, K.~Nemoto, {\em Teleportation of composite systems for communication and information processing}, arXiv:0803.1342.

\bibitem{teleport} C.H.~Bennett, G.~Brassard, C.~Crepeau, R.~Jozsa, A.~Peres, W.K.~Wootters, \prl {\bf 70}, 1895 (1993).

\bibitem{densecode} C.H.~Bennett and S.J.~Wiesner, \prl {\bf 69}, 2881 (1992).

\bibitem{Gn} A.~Sen(De), U.~Sen, and M.~Zukowski, \pra {\bf 68}, 032309 (2003).

\bibitem{murao} M.~Murao, D.~Jonathan, M.B.~Plenio, V.~Vedral, \pra {\bf 59}, 156 (1999).

\bibitem{D4.2} N.~Kiesel, C.~Schmid, G.~T\'oth, E.~Solano, and H.~Weinfurter, \prl {\bf 98}, 063604 (2007).

\bibitem{knill} E.~Knill, {\em Fermionic Linear Optics and Matchgates}, quant-ph/0108033.

\bibitem{terhal} B.M.~Terhal and D.P.~DiVincenzo, \pra {\bf 65}, 032325 (2002).

\bibitem{engel} H.-A.~Engel and D.~Loss, Science {\bf 309}, 586 (2005).

\bibitem{kolli} A.~Kolli, B.W.~Lovett, S.C.~Benjamin, and T.M.~Stace, \prl {\bf 97}, 250504 (2006); quant-ph/0607028. %% All-Optical Measurement Based QIP in Quantum Dots

\bibitem{coish} W.A.~Coish, V.N.~Golovach, J.C.~Egues, and D.~Loss, physica status solidi (b) {\bf 243}, 3658 (2006); cond-mat/0606782. %% Measurement, control, and decay of quantum-dot spins

\bibitem{charge_parity} B.~Trauzettel, A.N.~Jordan, C.W.J.~Beenakker, and M.~B\"uttiker, \prb {\bf 73}, 235331 (2006);  cond-mat/0602514.

\bibitem{stace1} T.M.~Stace, S.D.~Barrett, H.-S.~Goan, G.J.~Milburn, \prb {\bf 70}, 205342 (2004); cond-mat/0410181.

\bibitem{virmani} T.~Rudolph and S.S~Virmani, New J.~Phys.~{\bf 7}, 228 (2005).

\bibitem{mao} W.~Mao, D.V.~Averin, R.~Ruskov, and A.N.~Korotkov, \prl {\bf 93}, 056803 (2004); cond-mat/0401484.

\bibitem{ruskov} R.~Ruskov and A.N.~Korotkov, \prb {\bf 67}, 241305(R) (2003); cond-mat/0206396.

\bibitem{nemoto} K.~Nemoto and W.J.~Munro, \prl {\bf 93}, 250502 (2004).

\bibitem{munro} W.J.~Munro, K.~Nemoto, T.P.~Spiller, New J.~Phys.~{\bf 7}, 137 (2005).

\bibitem{spiller} T.P.~Spiller, K.~Nemoto, S.L.~Braunstein, W.J.~Munro, P.~van Loock, G.J.~Milburn, New J.~Phys.~{\bf 8}, 30 (2006).

\bibitem{nv1} A.D.~Greentree, J.~Salzman, S.~Prawer, and L.C.L.~Hollenberg, \pra {\bf 73}, 013818 (2006).

\bibitem{ph_module2} A.M.~Stephens, Z.W.E.~Evans, S.J.~Devitt, A.D.~Greentree, A.G.~Fowler, W.J.Munro, J.L.O'Brien, K.~Nemoto, L.C.L.Hollenberg, \pra {\bf 78}, 032318 (2008).

\bibitem{tashima} T.~Tashima, S.K.~Ozdemir, T.~Yamamoto, M.~Koashi, N.~Imoto, \pra {\bf 77}, 030302 (2008).

\bibitem{w_states} W.~D\"ur, G.~Vidal, and J.I.~Cirac, \pra {\bf 62}, 062314 (2000).

\bibitem{dicke} R.H.~Dicke, Phys.~Rev.~{\bf 93}, 99 (1954).

\bibitem{toth} G.~T\'oth, \josa B {\bf 24}, 275 (2007).

\bibitem{guhne} O.~G\"uhne, F.~Bodoky, M.~Blaauboer, {\em Multiparticle entanglement under the influence of decoherence}, arXiv:0805.2873.

\bibitem{sen} A.~Sen(De) U.~Sen, M.~Wiesniak, D.~Kaszlikowski, and M.~Zukowski, \pra {\bf 68}, 062306 (2003).

\bibitem{W8} H.~H\"affner {\it et al.}, Nature {\bf 438}, 643 (2005).

\bibitem{GHZ6} D.~Leibfried {\it et al.}, Nature {\bf 438}, 639 (2005).


\end{thebibliography}
\end{document}